\documentstyle[aps,prb,epsf]{revtex}

\begin{document}

\twocolumn[\hsize\textwidth\columnwidth\hsize\csname
@twocolumnfalse\endcsname

\title{ Spin wave analysis to the spatially-anisotropic Heisenberg
antiferromagnet on  triangular lattice}

\author{Adolfo E. Trumper}

\address{The Abdus Salam International Centre for Theoretical Physics\\
Strada Costiera 11, P. O. Box 586, 34100 Trieste, Italy}

\maketitle

\begin{abstract}
We study the phase diagram at $T=0$ of  the antiferromagnetic
Heisenberg model on the triangular lattice with spatially-anisotropic
interactions. For  values of the anisotropy very
close to $J_{\alpha}/J_{\beta}=0.5$, conventional
spin wave theory
predicts that quantum fluctuations melt 
the classical structures, for $S=1/2$. For the regime
$J_{\beta}< J_{\alpha}$, it is shown that 
the incommensurate spiral phases survive until  
 $J_{\beta}/J_{\alpha}=0.27$, leaving a wide region
where the ground state is disordered. 
The existence of such nonmagnetic states suggests
the possibility of spin liquid behavior for intermediate values 
of the anisotropy.\\ 
\end{abstract}

\vskip2pc]

For a long time  frustrated quantum antiferromagnets 
have been intensively studied. In this context, the antiferromagnetic
Heisenberg model 
on a triangular lattice is a prototype for such systems.
From the proposition of Anderson and Fazekas that this  
model is a candidate to exhibit spin liquid behavior\cite{FA}, 
a lot of work was done to 
understand  the nature of its  ground state. Although there is 
a general conviction that the ground state is ordered with a magnetic 
vector ${\bf Q}=(4\pi/3,0)$\cite{orden,triang}, some
authors found a
situation very close 
to a critical one or no magnetic order at all, leaving the answer still 
controversial\cite{desorden,Leung}. A systematic way to study the role of
frustration
is to vary  the strength of the exchange interaction along 
 the bonds. Recently Bhaumik {\it et al.}\cite{bose}  explored 
the existence of collinear phases on triangular and pentagonal
lattices and proposed that the critical value of the anisotropy, 
below which the ground state has collinear order, can be taken as a
measure of frustration.  

From the experimental point of view, the unconventional properties of     
the organic superconductors $\kappa -(BEDT-TTF)_2X$
and their similarities with the cuprates\cite{science} renewed
the interest in 
the triangular topology. In particular, it was argued\cite{kino,kenzie}
 that the Hubbard model on a
triangular lattice with anisotropic interactions at half filling      
could be a good candidate to explain such  properties.
In the limit of strong coupling
this model can be mapped to the Heisenberg model
with anisotropic interactions $J_{\alpha}=t^2_{\alpha}/U$,
$J_{\beta}=t^2_{\beta}/U$ where $t_{\alpha}$
and $t_{\beta}$ are the anisotropic hoppings. Furthermore, experiments  
suggest that
the relevant values of $J_{\alpha}/J_{\beta}$ are about $0.3-1$ (see for
details Ref.\cite{kenzie}), so the
combined
effect of anisotropy and frustration will take an important role in these
materials. 

In this paper we address the phase diagram 
of the Heisenberg model on the triangular lattice
with spatially-anisotropic interactions by mean of  conventional spin
wave theory. Our approach
provides the values of anisotropy where nonmagnetic states appear 
signaling the possible existence of spin liquid  behavior. 

The Hamiltonian is: \\

\begin{equation}
\label{H} H= J_{\alpha}\sum_{\bf r,\delta_{\alpha}}
 {\bf S}_{{\bf r}}\cdot {\bf S}_{{\bf r}+{\bf \delta}_{\alpha}}
+J_{\beta}\sum_{\bf r,\delta_{\beta}}       
 {\bf S}_{{\bf r}}\cdot {\bf S}_{{\bf r}+{\bf \delta_{\beta}}}
\end{equation}

\noindent where  $J_{\alpha}$ and $J_{\beta}$ are positive and correspond
to 
 interactions along  directions $\delta_{\alpha}$ and $\delta_{\beta}$
respectively (see figure 1).

\begin{figure}
\narrowtext
\begin{centering}
\epsfxsize=1.7truein
\vbox{\hskip 0.05truein
\epsffile{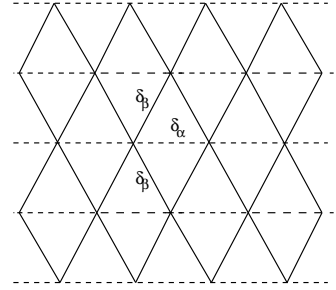}}
\medskip
\caption{Structure of the anisotropic bonds in a triangular lattice.}
\end{centering}
\label{}
\end{figure}

In order to develop a linear spin wave
theory, we need to know previously the classical
phase diagram. Basically, we replace the spin operators
by classical vectors on the $x-y$ plane and minimize the energy which is
equivalent to find 
the magnetic vector ${\bf Q}$ satisfying:\\
$$
J_{\bf Q}\leq J_{\bf k}, \forall\,\,\, {\bf k}
$$
where
\begin{equation}
\label{jk} J_{\bf k}=J_{\alpha}\,cos(k_x)+J_{\beta}\,2\,cos(\frac {k_x}
{2})\,cos(\frac
{k_y\sqrt3}
{2})
\end{equation}
The minimization of eq.(\ref{jk}) can be carried on easily for each value
of
$\mu = J_{\alpha}/J_{\beta}$ and it can be shown that
there are two kinds of  phases:\\
\begin{itemize}
\item Collinear: this state is characterized by 
${\bf Q}_{col}=(0,2\pi/\sqrt 3)$, and it is stable in the region
$0\leq \mu \leq 0.5$. The case $\mu=0$
is
topologically equivalent
to 
a square lattice and ${\bf Q}_{col}$ on a triangular lattice produces the
same magnetic structure than  $(\pi,\pi)$ on a square one. \\
\item  Incommensurate spiral:
 in this state ${\bf Q}_{spi}=(2Q,0)$, where 
$Q=cos^{-1}(-1/2\mu)$, and it is stable in the
region $0.5< \mu<\infty$.
For  $\mu=1$
we have the pure frustrated case which corresponds to the $120^{\circ}$ 
commensurate
spiral order and for  $\mu=\infty$ we have  infinite
decoupled classical chains each one N\'eel  ordered. 
\end{itemize}

\noindent In figure 2 we represent the possible values of $\pm {\bf Q}$ in
the
Brillouin zone for different values of $\mu$. Using the
invariance in {\bf k}-space under translations 
${\bf G}=(\pm 2\pi,\mp 2\pi/\sqrt 3)$ we can see that the
transition between all the possible states is continuous.

\begin{figure}
\narrowtext
\begin{centering}
\epsfxsize=2truein
\vbox{\hskip 0.05truein
\epsffile{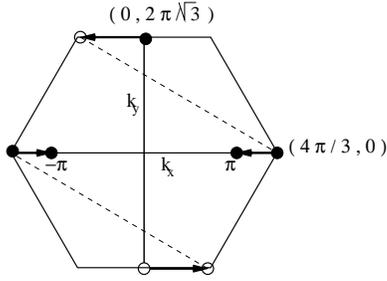}}
\medskip
\caption{ Magnetic vectors $\pm {\bf Q}$ for different
values of $\mu$ starting from  $\mu=0$
with $(0,2\pi/\sqrt 3)$ and ending at $\mu=\infty$ with $(\pi,0)$, in the
Brillouin zone. The equivalence between
empty and filled circles, represented by dashed lines, shows the
continuity of the classical transitions}
\end{centering}
\label{}
\end{figure}

We are
interested in how the transition between these classical sates is
affected by quantum fluctuations.
The strategy to perform the spin wave calculation is the one-sublattice
description. We apply a uniform twist of the coordinate frame in such
a way that x-axis direction coincides, in each site, with the
direction
of
the classical structure.
This allows us to incorporate quantum fluctuations in  a unique
way for  collinear and spiral phases. The next steps
are well known\cite{joliquer} and  we  only   describe
the procedure. i) The 
angular momentum operators are expressed by mean of the Holstein-Primakov
transformation, ii) the Hamiltonian is expanded to order $1/S$ (quadratic
order in bosons), iii) after Fourier transforming, the 
Hamiltonian can be diagonalized by a Bogoliubov transformation resulting: 
$$
H=E_c+\frac 12 \sum_{{\bf k}} (E({\bf k})-\gamma_{A}({\bf k}))+
  \frac 12 \sum_{{\bf k}} E({\bf k})\,\, (\alpha^{\dagger}_{{\bf k}}
   \alpha_{{\bf k}}+\alpha^{\dagger}_{{\bf -k}}\alpha_{{\bf -k}})
$$
where
$$
E_c=N\,S^2\,[J_{\beta}\,cos({\bf
Q}. {\bf \delta}_{\beta})+J_{\alpha}\,cos({\bf Q}. {\bf
\delta}_{\alpha})+J_{\beta}\,cos({\bf Q}. {\bf \delta}_{\beta})]
$$
and\\
\begin{equation}
\label{rel}E({\bf k})=[\gamma^{2}_{A}({\bf k})-\gamma^{2}_{B}({\bf
k})]^{-\frac 12}
\end{equation}
with
$$
\gamma_{A}({\bf k}))=\frac S2 \sum_{\delta=\delta_{\alpha},\delta_{\beta}} 
     J_{\delta}\,\, cos({\bf k}. {\bf \delta} )\,\, [1+cos({\bf Q}.
{\bf \delta} )]
     -2\,\, cos({\bf Q}. {\bf \delta} )
$$

$$
     \gamma_{B}({\bf k})=  S
      \sum_{\delta=\delta_{\alpha},\delta_{\beta}} 
      J_{\delta}\,\, cos(\,{\bf k}. {\bf \delta})\,\,
       [cos({\bf Q}.{\bf \delta} )-1]
$$
The compact equation (\ref{rel}) gives the dispersion relation
for
all the different phases labeled by ${\bf Q}$. In particular, for $\mu=0$ 
we recover the spin wave spectrum of the square lattice while for
$\mu=1$ we obtain the triangular one (spatially
isotropic). Independently of the value of ${\bf Q}$ it can be checked that 
$E({\bf k})=0$ for ${\bf k}={\bf 0},\pm{\bf Q}$, and these zero-modes
are  the
three
Goldstone modes related to  the complete symmetry-breaking of the $SU(2)$
invariance. However, for collinear phase ${\bf Q}$ is equivalent to $-{\bf
Q}$ and two zero-modes are recovered.
Maybe the most interesting result is that if  we expand  $E({\bf
k})$ near these zeros
 the behavior of $E({\bf k})$ is linear for all
$\mu\neq 1/2$, while
for $\mu= 1/2$, around ${\bf k}=(0,0)$ and along the direction of $k_x$ it
becomes quadratic showing the
softening of the spin wave modes for all $S$ (see figure 3).

\begin{figure}
\narrowtext
\begin{centering}
\epsfxsize=3.3truein
\vbox{\hskip 0.05truein
\epsffile{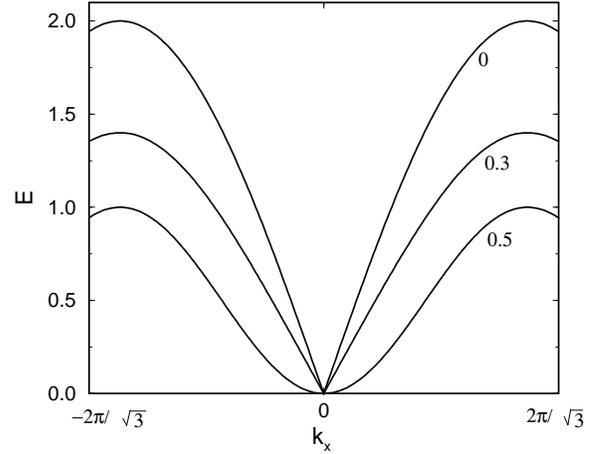}}
\medskip
\caption{Relation dispersion $E({\bf k})$ along $k_x$ direction for
different values of $\mu=J_{\alpha}/J_{\beta}$ and $S= 1/2$.}
\end{centering}
\label{}
\end{figure}

This is manifested in the 
magnetization,

\begin{equation}
\label{m0} m_0= (S+\frac 12) 
-\frac {\sqrt3}{2} \int_{BZ} \frac {dk_x dk_y}{(2\pi)^2}\frac
{\gamma_{A}({\bf k}))}{2\,E({\bf k})}  
\end{equation}

\noindent where the integration  is over the Brillouin zone of the
triangular lattice. In the second term of eq.(\ref{m0})   the 
integrand diverge for ${\bf k}={\bf
0},\pm{\bf Q}$ but the integral remains  finite in 2D for all
$\mu\neq 1/2$,
while for $\mu=1/2$ quantum fluctuations are amplified,
because of 
the quadratic behavior of $E({\bf k})$ near ${\bf k}=(0,0)$, 
and it can be proved\cite{Merino} that the integral 
contribute with a finite but large enhacement of quantum correction
to the magnetization. In order to estimate 
the region where the system is disordered for $S=1/2$,   
we calculated the quantum corrected magnetization for each classical
structure. In what follows we rescale $\mu \rightarrow
\eta=\mu/(1+\mu)$.
This allows to capture all the possible values of $\mu$ in a finite
range $0\leq \eta \leq 1$.

\begin{figure}
\narrowtext
\epsfxsize=3.3truein
\vbox{\hskip 0.05truein
\epsffile{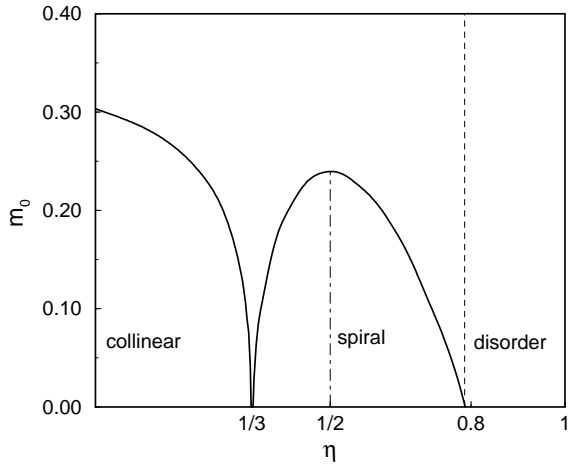}} 
\medskip
\caption{Quantum corrected magnetization Vs. $\eta$ for spin
$S=1/2$. Dot-dashed line indicates the location of commensurate
N\'eel order.}
\label{}
\end{figure}

\noindent Figure 4 shows that
 spin wave
theory predicts a
collinear
phase at $\eta =0$ with $m_0=0.303$ and as we increase the
frustration it is weaken continuously getting  disordered just before the
classical value $\eta=1/3$. Immediately after 
this value, an incommensurate order is stabilized starting from $m_0=0$
and it becomes  more robust as we approach the spatially isotropic case
$\eta=0.5$ where the structure is commensurate, ${\bf Q=}
(4\pi/3,0)$, and $m_0=0.239$. If we continue increasing  $\eta$ 
incommensurate structures appear again with decreasing $m_0$ until the
critical value $\eta= 0.79$ where magnetization vanishes. Beyond
this
value, the ground state is disordered. 
We  note that the singular behavior 
obtained near $\eta= 1/3$ does not appear in previous studies of 
this model. In
fact, Gazza {\it et al.}\cite{SP} 
performed a Schwinger boson mean field theory and  found a continuous 
transition from collinear to spiral phases at $\eta=0.375$ but
with a nonvanishing magnetization $m_0\sim 0.175$.  However,  inclusion
of gaussian fluctuations in this theory would tend to decrease the
order, as it is
known to occur in highly frustrated cases\cite{letter,triang},
reaching probably more accord with
our spin wave results. The same happens with both theories in
the $J_1-J_2$ model on a square lattice\cite{chandra,mila,letter}.
One should take into account  that our system can be thought 
as a Heisenberg model on a square lattice with interactions to first
and second neighbours, but only along one of the diagonals \cite{Zheng}.

On the other hand, for the regime $J_{\beta} < J_{\alpha}$, 
the critical value $\eta= 0.79$ means that  the system disorders 
at $J_{\beta}/J_{\alpha}= 0.27$. This should be compared with 
the spin wave value for the square case\cite{sandro}, $J_{\bot}/J_{\Vert} \sim
0.03$, where the difference in one order of magnitude
shows that the way in which fluctuations overcome the ordering
is different. However, in the regime around $\eta =1$ the spin wave 
calculation is not reliable any more since quantum fluctuations
are divergent in the 1D limit. In particular,  numerical  techniques\cite{cuad} 
 predict that  in the square case
an infinitesimal coupling is
required  to take the chains away from criticality and get ordered.
A similar calculation should be done for our model in the regime 
of weakly coupled chains and it is left for a future work.

In conclusion, we have studied the Heisenberg model on a triangular
lattice 
with spatially-anisotropic interactions by mean of a spin wave analysis.
We calculated the classical and quantum corrected phase diagram at $T=0$ 
for the whole range of parameters
$\eta=J_{\alpha}/(J_{\alpha}+J_{\beta})$ obtaining
different regimes: collinear, incommensurate spirals and disorder
phases. The nonmagnetic region found
very near  
 the singular value $\eta=1/3$  for  $S=1/2$ suggests  the
possible existence of a spin liquid phase. A similar scenario occurs
in the $J_1-J_2$ and $J_1-J_3$ model on a square
lattice\cite{chandra,locher}. It is
clear from our
approximation that it should
be more probable to find a spin liquid behavior near $\eta=1/3$
than
in other region of the diagram between collinear and spiral phases.
Though this region is small,
it is  just located in the range where  the experimental  values of
$J_{\alpha}/J_{\beta}$     
are relevant for organic superconductors $\kappa -(BEDT-TTF)_2X$.
Moreover, in the regime of $J_{\beta}<J_{\alpha}$
we found that quantum fluctuations destroy the order at $J_{\beta}/J_{\alpha}=0.27 $ 
leaving a wide region where the ground state is disordered.

Finally, we would like to stress  that by applying a  simple
approximation like spin wave theory we have obtained a very rich phase 
diagram. Of course, we have
not demonstrated the existence of spin liquid
phases but the appearance of
nonmagnetic regions indicates the possible location of them.
Another quantities
like spin gap or correlation functions are needed to explore more deeply 
the nature of these phases, and it requires  more powerful techniques.\\

The author acknowledges the useful discussions with A. Ceccatto,
S. Sorella and E. Jagla. This work was supported by Consejo
Nacional de Investigaciones Cient\'{\i}ficas y T\'ecnicas of Argentina.\\

{\it Note added}: during revision of this work we became aware of two recent 
works performed on this model. The first one by Zheng {\it et al.} \cite{Zheng} using
series expansion technique where their prediction are, in general, similar to the phase
diagram obtained in this work. The second one, Merino {\it et al.}\cite{Merino} using the 
same technique of the present work.

\end{document}